\documentclass[10pt, twocolumn, secnumarabic, amssymb, nobibnotes, footinbib, showpacs, aps, pra, superscriptaddress]{revtex4-1}

\usepackage{graphicx}
\usepackage{dcolumn}
\usepackage{amsmath}
\usepackage{mathtools}
\usepackage{natbib}
\usepackage{color}
\usepackage{epstopdf}

\begin{document}

\title{Nonlocal bunching of composite bosons}

\author{Zakarya Lasmar}
\affiliation{Faculty of Physics, Adam Mickiewicz University, Umultowska 85, 61-614 Pozna\'n, Poland}
\author{Dagomir Kaszlikowski}    
\affiliation{Centre for Quantum Technologies,
National University of Singapore, 3 Science Drive 2, 117543 Singapore,
Singapore}
\affiliation{Department of Physics,
National University of Singapore, 3 Science Drive 2, 117543 Singapore,
Singapore}
\author{Pawe\l{} Kurzy\'nski}   \email{pawel.kurzynski@amu.edu.pl}  
\affiliation{Faculty of Physics, Adam Mickiewicz University, Umultowska 85, 61-614 Pozna\'n, Poland}
\affiliation{Centre for Quantum Technologies, National University of Singapore, 3 Science Drive 2, 117543 Singapore, Singapore}

\date{\today}

%%%%%%%%%%%%%%%%%%%%%%%%%%%%%%%%%%%%%%%%%%%%%%%%%%%

\begin{abstract}
It was suggested that two entangled fermions can behave like a single boson and that the bosonic quality is proportional to the degree of entanglement between the two particles. The relation between bosonic quality and entanglement is quite natural if one takes into account the fact that entanglement appears in bound states of interacting systems. However, entanglement can still be present in spatially separated subsystems that do not interact anymore. These systems are often a subject of studies on quantum nonlocality and foundations of quantum physics. Here, we ask whether an entangled spatially separated fermionic pair can exhibit bosonic properties. We show that in certain conditions the answer to this question can be positive. In particular, we propose a nonlocal bunching scenario in which two such pairs form an analogue of a two-partite bosonic Fock state. 
\end{abstract}

%%%%%%%%%%%%%%%%%%%%%%%%%%%%%%%%%%%%%%%%%%%%%%%%%%%


\maketitle

%%%%%%%%%%%%%%%%%%%%%%%%%%%%%%%%%%%%%%%%%%%%%%%%%%%

\section{Introduction}

As far as we know, elementary particles can either be bosons or fermions. For instance, photons and electrons are particles that exhibit the ideal bosonic and fermionic behaviour, respectively. Apart from restrictions on occupation numbers, which are expressed by the Pauli exclusion principle, the simplest method to distinguish between fermions and bosons is via the celebrated Hong-Ou-Mandel two-particle interference \cite{HOM}. If two bosons meet on a symmetric beam splitter, they bunch -- they come out together through the same port. On the other hand, two fermions come out separately, i.e., they anti-bunch. 

In real world most of particles that we encounter are not elementary. Atoms, molecules, or even protons and neutrons are composed of a few elementary particles. Majority of bosons are in fact made of an even number of elementary fermions. The constituents of such composite bosons (cobosons) are bounded via strong forces that keep them together and constrain most of degrees of freedom such that the total system is effectively described by the centre of mass and total momentum. Therefore, it is legitimate to threat such systems as a single particle as long as forces acting on it are weaker than the binding forces. 

Interestingly, a multipartite bound state is often highly entangled. This is clearly visible for cobosons made of two components, such as hydrogen or exciton, for which the total state is pure, but the state of each subsystem is highly mixed. It was proposed that the bosonic quality of such particles is proportional to the degree of entanglement between the constituents \cite{Law}. This idea was further developed in a number of works \cite{Chudzicki, Law,e2,e3,e4,e5,e6,e7,e8,e9,e10,e11,e12,e13,e14,e15,e16,e17,e18,e19,e20,e21,e22}. Since intra-particle entanglement can exist even if particles are spatially separated and there is no interaction between them, it is natural to ask if two spatially separated entangled fermions can still manifest some kind of bosonic behaviour. This is the problem we consider in this work. 

More precisely, we discuss the stability of a coboson undergoing the beam splitter transformation. We show that stability requires entanglement production, therefore some kind of interaction is required in order to keep the coboson intact. Next, we study the problem of two-partite bunching on a symmetric beam splitter. In case of elementary bosons the initial state $a_L^{\dagger}a_R^{\dagger}|0\rangle\equiv |1\rangle|1\rangle$, where $a^{\dagger}_L$ and $a^{\dagger}_R$ create particles in modes $L$ and $R$ respectively, is transformed into $(a_L^{\dagger 2}+a_R^{\dagger 2})/2|0\rangle \equiv 1/\sqrt{2}(|2\rangle|0\rangle + |0\rangle|2\rangle)$. In case of composite bosons both particles have internal structure and we study the entanglement properties of the corresponding states. We show, that differences in entanglement in the initial and final bipartite cobosonic states are much more subtle than in case of single-partite cobosonic states. Finally, we propose a nonlocal bunching scenario in which the interaction is only between the local parts and show that the probability of perfect bunching is proportional to the degree of entanglement between the constituents, which is in accordance with previous results. 

%%%%%%%%%%%%%%%%%%%%%%%%%%%%%%%%%%%%%%%%%%%%%%%%%%%%%%%%%%%%%%%%%%%%	
		
\section{Preliminaries}

\subsection{Coboson made of two fermions}

Let us consider two distinguishable fermions, say electron and hole, whose creation operators are denoted by $a_m^\dagger$ and $b_n^{\dagger}$. Each particle can occupy $d$ different modes, therefore $m,n=1,\dots,d$. The general state of these two particles is given by
\begin{equation}\label{s1}
|\psi\rangle = \sum_{m,n=1}^d \gamma_{m,n} a_m^{\dagger}b_n^{\dagger}|0\rangle.
\end{equation}
However, one can find a basis transformations $a_i^{\dagger} = \sum_m \alpha_{i,m}a_m^{\dagger}$ and $b_i^{\dagger} = \sum_n \beta_{i,n}b_n^{\dagger}$ such that the state (\ref{s1}) becomes
\begin{equation}\label{s2}
|\psi\rangle = \sum_{i=1}^d \sqrt{\lambda_i} a_i^{\dagger}b_i^{\dagger}|0\rangle.
\end{equation}
Such transformation is known as Schmidt decomposition \cite{book} and is very helpful to determine the degree of entanglement between particle $a$ and $b$. One can always find a set of parameters $\{\lambda_i\}_{i=1}^{d}$ such that $\lambda_i \geq 0$ and $\lambda_{i+1} \geq \lambda_i$. The number of nonzero $\lambda_i$'s is known as a Schmidt rank and the system is entangled whenever the Schmidt rank is greater than one. 

Since the state of coboson is pure, the degree of entanglement can be measured by purity 
\begin{equation}
P=\sum_{i=1}^d \lambda_i^2,
\end{equation}
which corresponds to $P=\text{Tr}\{\rho_A^2\}=\text{Tr}\{\rho_B^2\}$, where $\rho_A$ and $\rho_B$ are density matrices of particles $a$ and $b$, respectively. Such density matrices can be calculated from the joint state $|\psi\rangle$ using the method discussed in \cite{Yang}. For example, the matrix element $\rho_{A_{n,m}}$ corresponds to
\begin{equation}\label{rho}
\rho_{A_{n,m}} = \langle \psi|a_m^{\dagger}a_n |\psi\rangle.
\end{equation}
Analogous method applies to $\rho_B$. Therefore
\begin{eqnarray}
\rho_{A} &=& \sum_{i=1}^d \lambda_i a_i^{\dagger} |0\rangle\langle 0|a_i,~~ \\
\rho_{B} &=& \sum_{i=1}^d \lambda_i b_i^{\dagger} |0\rangle\langle 0|b_i.
\end{eqnarray}
The purity is bounded by $\frac{1}{d} \leq P \leq 1$ and the smaller the purity, the more entangled the system is. For $P=1$ the system is separable. 

Next, we associate $|\psi\rangle$ with a single particle state 
\begin{equation}
|\psi\rangle=c^{\dagger}|0\rangle\equiv  |1\rangle,
\end{equation}
where 
\begin{equation}\label{coperator}
c^{\dagger}= \sum_{i=1}^d \sqrt{\lambda_i} a_i^{\dagger}b_i^{\dagger}
\end{equation}
is a creation operator of a single coboson. The modes labeled by $i$ can be considered as an internal structure of a coboson. 

The properties of the operator $c^{\dagger}$ were extensively studied by many researchers, for a review see \cite{Combescot}. In particular, the purity resulting from state $|\psi\rangle$ was associated with the quality of cobosonic creation and annihilation \cite{Law}. Consider a Fock state of $N \leq d$ composite particles
\begin{equation}
|N\rangle \equiv \chi_N^{-1/2}\frac{c^{\dagger N}}{\sqrt{N!}}|0\rangle,
\end{equation}
where $\chi_N$ is a normalisation factor such that $\langle1|1\rangle =1$ 
\begin{equation}
\chi_N=N!\sum_{n_1 < \dots < n_N}\lambda_{n_1}\dots \lambda_{n_N}.
\end{equation}
We get
\begin{eqnarray}
&c^\dagger& |N-1\rangle = \alpha_N \sqrt{N}|N\rangle, \\
&c& |N\rangle = \alpha_N \sqrt{N} |N-1\rangle + |\epsilon_N\rangle,
\end{eqnarray}
where
\begin{equation}
\alpha_N= \sqrt{\frac{\chi_N}{\chi_{N-1}}},
\end{equation}
and $|\epsilon_N\rangle$ is a state of $N-1$ pairs of particles $a$ and $b$ which do not correspond to a $N-1$-partite coboson Fock state. In simple words, $|\epsilon_N\rangle$ corresponds to a state in which coboson structure is destroyed. The norm of this state is given by
\begin{equation}
\langle \epsilon_N|\epsilon_N\rangle = 1 - N\frac{\chi_N}{\chi_{N-1}}+(N-1)\frac{\chi_{N+1}}{\chi_{N}}.
\end{equation}

We see that the bosonic ladder structure is recovered when $\alpha_N \rightarrow 1$ and $\langle \epsilon_N|\epsilon_N\rangle \rightarrow 0$. This happens when $\frac{\chi_{N+1}}{\chi_{N}} \rightarrow 1$ for all $N$. It was shown in \cite{Law,Chudzicki} that this ratio is bounded by purity
\begin{equation}
1-NP \leq \frac{\chi_{N+1}}{\chi_{N}} \leq 1-P.
\end{equation}
Therefore, in the limit of maximal entanglement $P=\frac{1}{d}$ one has
\begin{equation}
1-\frac{N}{d} \leq \frac{\chi_{N+1}}{\chi_{N}} \leq 1-\frac{1}{d}. 
\end{equation}
As a result, $\frac{\chi_{N+1}}{\chi_{N}} \rightarrow 1$ once $N/d \ll 1$.

%%%%%%%%%%%%%%%%%%%%%

\subsection{Maximally entangled cobosons}

The cobosonic creation operator defined in Eq. (\ref{coperator}) is of the most general form. The entanglement between two fermions depends on a choice of parameters $\lambda_i$. Here, we choose a simplified version of $c^{\dagger}$ which is only parametrised by $d$, i.e., the dimension of a single-particle Hilbert space. Nevertheless, it is straightforward to show that the results of this work will also hold if one chooses general operators. The corresponding coboson is a maximally entangled state of two qudits
\begin{equation}\label{c}
 c_{}^\dag = \frac{1}{\sqrt{d}} \sum_{i = 1 }^{d}  a_{i}^\dag b_{i}^\dag .
\end{equation}
In this case 
\begin{eqnarray}
& &\chi_N = \frac{d!}{d^N(d-N)!}, \\
& &\alpha_N = \sqrt{\frac{d-N+1}{d}}, \\
& &\langle \epsilon_N|\epsilon_N\rangle = 0;
\end{eqnarray}
Because of the last equality the ladder structure of creation and annihilation operators is recovered, however it deviates from perfect bosonic structure due to factors $\alpha_N$.

%%%%%%%%%%%%%%%%%%%%%

\section{Entanglement and stability under beam splitting transformation}

\subsection{Single coboson}

\begin{figure}[t]
\includegraphics[width=0.45 \textwidth,trim=4 4 4 4,clip]{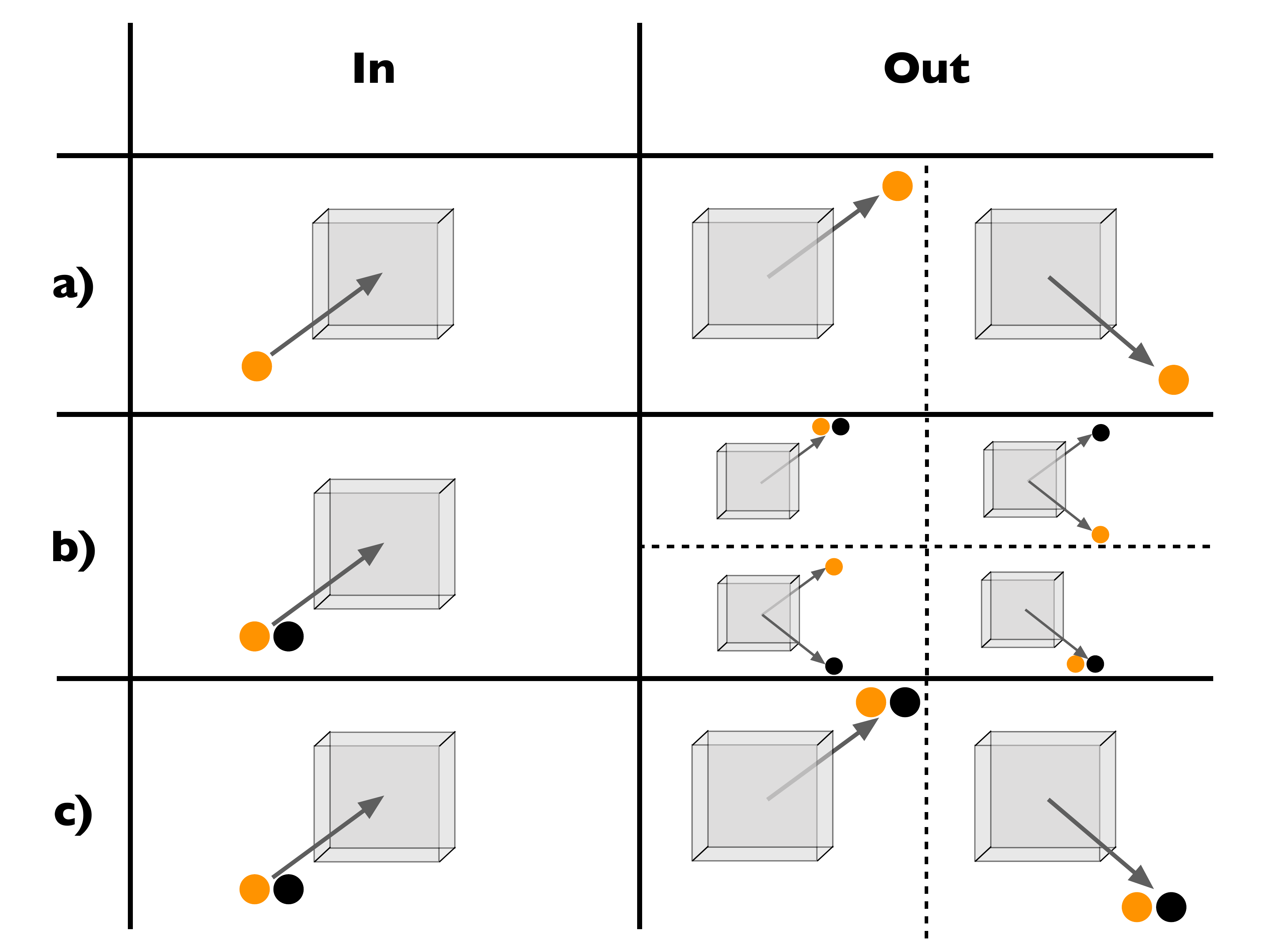}
\caption{Beam-splitting of elementary and composite particles. a) A single elementary particle can either go through or reflect from BS. b) Two non-interacting particles evolve independently and each of them can either go through or reflect from BS, which leads to four possible outcomes. If the input state is considered to represent a single composite particle, then the evolution inevitably leads to its decay. c) Two interacting particles stay together - they collectively go through or reflect from BS which allows to treat them as a single particle.}
\label{fig1}
\end{figure}

Let us consider a beam splitter (BS) whose Hamiltonian is given by
\begin{equation}\label{Hbs}
H_{BS}=a^{\dagger}_L a_R + a^{\dagger}_R a_L, 
\end{equation}
where $R$ and $L$ denote the right and the left BS mode. If the time of the evolution is $t=\frac{\pi}{4}$ we get a symmetric BS. In such case a single particle in the right mode undergoes a transformation
\begin{equation}
a_R^{\dagger}|0\rangle \rightarrow \frac{1}{\sqrt{2}}(a_R^{\dagger} -i a_L^{\dagger})|0\rangle.
\end{equation}
Similarly, a particle in the left mode transforms as 
\begin{equation}
a_L^{\dagger}|0\rangle \rightarrow \frac{1}{\sqrt{2}}(a_L^{\dagger} -i a_R^{\dagger})|0\rangle.
\end{equation}
Therefore, a single particle either goes through or is reflected from BS, see Fig. \ref{fig1} a).

%%%%%%%%%%%%%%%%%%%%%

%%%%%%%%%%%%%%%%%%%%%

Next, we consider two fermions in a cobosonic state (\ref{c}). This time particles are described by two degrees of freedom $a^{\dagger}_{i,X}$ and $b^{\dagger}_{i,Y}$, where $i=1,\dots,d$ and $X,Y=R,L$. Each particle evolves independently according to a modified version of (\ref{Hbs}) 
\begin{eqnarray}
H_A &=& \sum_{i=1}^d (a^{\dagger}_{i,L} a_{i,R} + a^{\dagger}_{i,R} a_{i,L}), \label{Ha}\\
H_B &=& \sum_{i=1}^d (b^{\dagger}_{i,L} b_{i,R} + b^{\dagger}_{i,R} b_{i,L}), \label{Hb}
\end{eqnarray}
which leads to
\begin{eqnarray}
& &c^{\dagger}_L|0\rangle \equiv \frac{1}{\sqrt{d}} \sum_{i = 1 }^{d}  a_{i,L}^\dag b_{i,L}^\dag |0\rangle \rightarrow \label{indep}\\
& &\frac{1}{\sqrt{2d}} \sum_{i = 1 }^{d}  (a_{i,L}^\dag b_{i,L}^\dag - ia_{i,R}^\dag b_{i,L}^\dag -i a_{i,L}^\dag b_{i,R}^\dag -a_{i,R}^\dag b_{i,R}^\dag) |0\rangle.  \nonumber
\end{eqnarray}
We see that the independent evolution leads to a decay of a composite boson, since in half of the cases the fermion $a$ will exit through a different port than the fermion $b$, see Fig. \ref{fig1} b).

Finally, let us discuss a BS transformation of a coboson whose components are interacting. This problem has been studied before, see for example \cite{e16,e17,e18,e22}, and we consider the interaction model similar to the one proposed in \cite{e17}. Apart from (\ref{Ha}) and (\ref{Hb}) the Hamiltonian contains an interaction term
\begin{equation}
H_{int}=-\gamma \sum_{X=R,L}\sum_{i=1}^d a_{i,X}^{\dagger}a_{i,X}b_{i,X}^{\dagger}b_{i,X}.
\end{equation}
In the limit $\gamma \gg 1$ the evolution of the system can be approximated as
\begin{eqnarray}
& &c^{\dagger}_L|0\rangle \equiv \frac{1}{\sqrt{d}} \sum_{i = 1 }^{d}  a_{i,L}^\dag b_{i,L}^\dag |0\rangle \rightarrow \label{int} \\
& &\frac{1}{\sqrt{2d}} \sum_{i = 1 }^{d}  (a_{i,L}^\dag b_{i,L}^\dag  - a_{i,R}^\dag b_{i,R}^\dag) |0\rangle \equiv \frac{1}{\sqrt{2}}(c^{\dagger}_L - c^{\dagger}_R)|0\rangle. \nonumber 
\end{eqnarray}
The above describes a collective behaviour of two fermions, they either both reflect or both go through BS. Therefore, the evolution can be interpreted as a single particle behaviour, because there is no way to detect any internal structure of the system, see Fig. \ref{fig1} c).

Now, let us think for a moment why interaction is so important for the stability of the system. First, we note that the BS model considered by us describes the evolution of only one degree of freedom $X=R,L$. The other degree of freedom, corresponding to the internal structure of the coboson, is decoupled from the evolution. Therefore, let us consider for a moment the evolution of an unentangled fermionic pair, since the intra-particle entanglement in operator (\ref{c}) seems to play no role in this case. At the moment, we say nothing about the details of the evolution, we just demand that the following transformation takes place
\begin{equation}\label{trans}
a_L^{\dagger}b_L^{\dagger} \rightarrow \frac{1}{\sqrt{2}}(a_L^{\dagger}b_L^{\dagger}+e^{i\varphi}a_R^{\dagger}b_R^{\dagger})|0\rangle,
\end{equation}
where $\varphi$ is an arbitrary phase. It resembles a single-particle transformation, however this time the final state of a single fermion is mixed, for example
\begin{equation}
\rho_A=\frac{1}{2}(a_L^{\dagger}|0\rangle\langle 0|a_L+a_R^{\dagger}|0\rangle\langle 0|a_R).
\end{equation}
On the other hand, the initial state of each fermion is pure. Therefore, the transformation (\ref{trans}) generates entanglement between the two fermions which can only happen if they interact either directly or via some mediating auxiliary system. This proves that stability of a single composite particle under BS transformation requires entanglement production, which implies that some kind of interaction is inevitable. Finally, note that entanglement production can be also observed in the transformation (\ref{int}). The purity of a single fermion changes from $\frac{1}{d}$ to $\frac{1}{2d}$. It can be also easily evaluated that in case of transformation (\ref{indep}) the purity does not change. 

%%%%%%%%%%%%%%%%%%%%%

\subsection{Two cobosons}

Next, we consider a transformation of two cobosons in a state $c^{\dag}_L c^{\dag}_R|0\rangle$. As we mentioned in the introduction, elementary bosons bunch, i.e., the initial state $a_L^{\dag}a_R^{\dag}|0\rangle$ transforms into $\frac{1}{2}(a_L^{\dag 2} + a_R^{\dag2})|0\rangle$. Now, we consider a transformation
\begin{eqnarray}
& &|\psi_i\rangle=c^{\dag}_L c^{\dag}_R|0\rangle \equiv \frac{1}{d}\sum_{i,j=1}^d a_{i,L}^\dag b_{i,L}^\dag a_{j,R}^\dag b_{j,R}^\dag |0\rangle \rightarrow \label{cbunch} \\
& &\frac{1}{d\sqrt{\chi_2}}\sum_{i,j=1}^d (a_{i,L}^\dag b_{i,L}^\dag a_{j,L}^\dag b_{j,L}^\dag + a_{i,R}^\dag b_{i,R}^\dag a_{j,R}^\dag b_{j,R}^\dag)|0\rangle  \nonumber \\
& &\equiv \frac{(c_L^{\dag 2} + c_R^{\dag2})}{2\sqrt{\chi_2}}|0\rangle=|\psi_f\rangle, \nonumber
\end{eqnarray}
where
\begin{equation}
\chi_2=\frac{d-1}{d}=1-P.
\end{equation}
Just like in the previous section, at the moment we are not interested what kind of evolution causes this transformation. Our current goal is to examine the entanglement properties of $|\psi_i\rangle$ and $|\psi_f\rangle$.

First, let us note that although each coboson is made of two distinguishable fermions, a two-cobosonic state is made of two fermions of type $a$ and two fermions of type $b$. Fermions of the same type are indistinguishable and therefore we need to be careful with how we define entanglement. In general, one should use the approach which was extensively discussed in \cite{Eckert}, however for our purposes it is enough to study purity of certain subsystems.

We start with a purity of a single particle. We choose particle $a$, however due to symmetry of the states the purity of $b$ is the same. We use Eq. (\ref{rho}) to determine $\rho_{A_i}$ and $\rho_{A_f}$. Note, that because right now there are two particles of the same type, the trace of both density matrices is two \cite{Yang}. After renormalisation we find
\begin{equation}
\rho_{A_i}=\rho_{A_f}=\frac{1}{2d}\sum_{X=L,R}\sum_{i=1}^d a_{i,X}^{\dagger}|0\rangle\langle 0|a_{i,X}.
\end{equation}
As we can see, the single-particle state does not change under the transformation (\ref{cbunch}). In addition, the corresponding purity is $\frac{1}{2d}$.

Next, we calculate a two-partite state of particles $a$. In this case matrix elements are given by
\begin{equation}
\rho_{A_{kl,nm}}=\langle\psi|a_m a_n a_k^{\dagger}a_l^{\dagger}|\psi\rangle.
\end{equation}
We get
\begin{eqnarray}
\rho_{A_i} &=& \frac{1}{d^2}\sum_{i,j=1}^d a_{i,L}^{\dagger}a_{j,R}^{\dagger}|0\rangle\langle 0|a_{j,R}a_{i,L},\\
\rho_{A_f} &=& \frac{1}{d^2 \chi_2} \sum_{\substack{i , j = 1 \\ i>j}}^d \left( a_{i,L}^{\dagger}a_{j,L}^{\dagger}|0\rangle\langle 0|a_{j,L}a_{i,L} \right. \\ 
 &+&  \left. a_{i,R}^{\dagger}a_{j,R}^{\dagger}|0\rangle\langle 0|a_{j,R}a_{i,R}\right). \nonumber
\end{eqnarray}
This time the initial state of two particles is different than the final one. Moreover, the initial two-particle purity is $P^{(2)}_i=\frac{1}{d^2}$ and the final one is $P^{(2)}_f=\frac{1}{d^2 \chi_2}=\frac{1}{d(d-1)}$.

We observe that the transformation (\ref{cbunch}) causes a change at a level of two particles of the same type, but not at the level of a single particle. Interestingly, unlike in the case of a single particle, the entanglement between particles of type $a$ and $b$ decreases since $P^{(2)}_i < P^{(2)}_f$. This suggests that the transformation (\ref{cbunch}) may not require interaction between particles of type $a$ and $b$, but rather some interaction between particles of the same type and perhaps some post-selection which would reduce the entanglement.

%%%%%%%%%%%%%%%%%%%%%

\section{Nonlocal bunching}

In order to confirm the above conjecture, one can consider a special scenario in which particles $a$ and $b$ are spatially separated, though still entangled. Note, that such scenario is also in accordance with our primary question: do spatially separated entangled fermions exhibit some kind of bosonic behaviour? 

We consider a typical Bell-like setup \cite{Bell}, however this time our goal is not to disprove local realistic description of measurements performed on spatially separated subsystems, but to show that the transformation (\ref{cbunch}) is realisable via solely local operations. We have two spatially separated experimenters, Alice and Bob, who share two cobosons. More precisely, each coboson is split into basic constituents and fermions of type $a$ go to Alice whereas fermions of type $b$ go to Bob. Note, that entanglement between $a$ and $b$ is still present, therefore formally the system is still described by the operator (\ref{c}). Moreover, we assume that constituents corresponding to different cobosons are initially occupying different modes, which we label $X=R,L$, just like in the previous section. Therefore, it is legitimate to describe the initial state as
\begin{equation}\label{psii}
|\psi_i\rangle=c^{\dag}_L c^{\dag}_R|0\rangle \equiv \frac{1}{d}\sum_{i,j=1}^d a_{i,L}^\dag b_{i,L}^\dag a_{j,R}^\dag b_{j,R}^\dag |0\rangle.
\end{equation}

%%%%%%%%%%%%%%%%%%%%%

\begin{figure}[t]
\includegraphics[width=0.45 \textwidth,trim=4 4 4 4,clip]{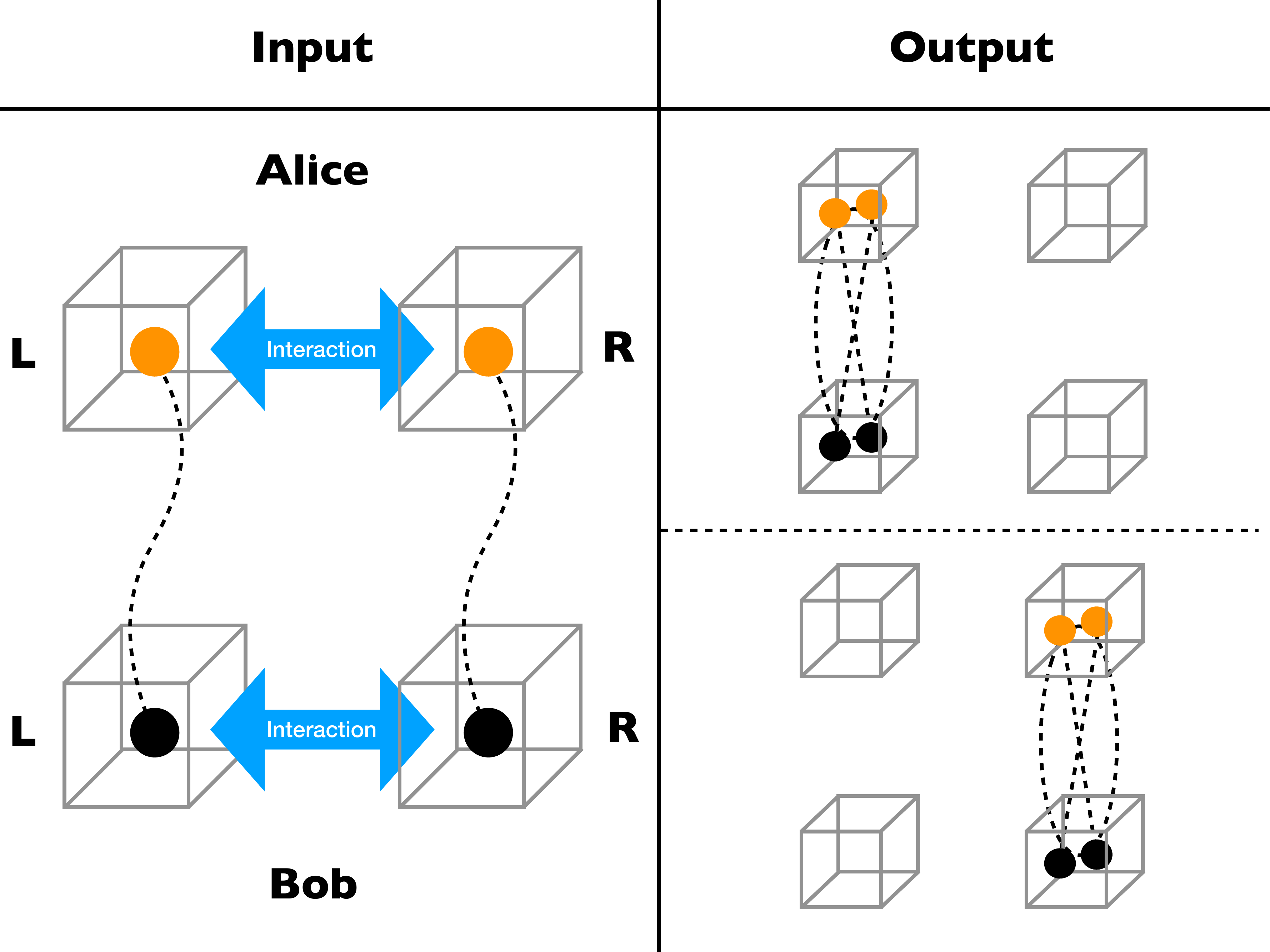}
\caption{Schematic representation of the nonlocal bunching of two cobosons. It is impossible to say which pairs of particles form a coboson once they are in the same mode.}
\label{fig2}
\end{figure}

%%%%%%%%%%%%%%%%%%%%%

Next, we allow particles of the same type to interact via the following Hamiltonians (see Fig. \ref{fig2}),
\begin{eqnarray} \label{eq: Ham}
{\cal H}_{A} = \sum_{\substack{i , j = 1 \\ i>j}}^{d} & \big( a_{i,L}^\dag a_{j,L}^\dag a_{j,R} a_{i,L} + a_{i,R}^\dag a_{j,R}^\dag a_{i,R} a_{j,L} \nonumber\\
+ & a_{i,L}^\dag a_{j,R}^\dag a_{j,L} a_{i,L} + a_{j,L}^\dag a_{i,R}^\dag a_{j,R} a_{i,R}\big),
\end{eqnarray}
and
\begin{eqnarray} \label{eq: Ham}
{\cal H}_{B} = \sum_{\substack{i , j = 1 \\ i>j}}^{d} & \big( b_{i,L}^\dag b_{j,L}^\dag b_{j,R} b_{i,L} + b_{i,R}^\dag b_{j,R}^\dag b_{i,R} b_{j,L} \nonumber\\
+ & b_{i,L}^\dag b_{j,R}^\dag b_{j,L} b_{i,L} + b_{j,L}^\dag b_{i,R}^\dag b_{j,R} b_{i,R}\big).
\end{eqnarray}
This is a local evolution, since particles of the same type are in the same spatial location. For $t=\frac{\pi}{2}$ the Hamiltonian ${\cal H}_{A}$ generates the following transformations
\begin{eqnarray}
& & a_{i,L}^{\dag}a_{j,R}^{\dag} \rightarrow -i a_{i,L}^{\dag}a_{j,L}^{\dag} ~~~~\text{for}~~ i>j, \\
& & a_{j,L}^{\dag}a_{i,R}^{\dag} \rightarrow -i a_{i,R}^{\dag}a_{j,R}^{\dag} ~~~~\text{for}~~ i>j, \\
& & a_{i,L}^{\dag}a_{i,R}^{\dag} \rightarrow  a_{i,L}^{\dag}a_{i,R}^{\dag}.
\end{eqnarray}
Analogous transformations are generated by ${\cal H}_{B}$. Therefore, the state (\ref{psii}) is transformed into
\begin{eqnarray} \label{psif}
& & \dfrac{1}{d} \Big( \sum_{i > j = 1 }^{d}  a_{i,\,L}^\dag b_{i,\,L}^\dag a_{j,\,L}^\dag b_{j,\,L}^\dag + a_{i,\,R}^\dag b_{i,\,R}^\dag a_{j,\,R}^\dag b_{j,\,R}^\dag  \nonumber\\
&-& \sum_{k = 1 }^{d}  a_{k,\,L}^\dag b_{k,\,L}^\dag a_{k,\,R}^\dag b_{k,\,R}^\dag \Big)  |0\rangle  \nonumber\\
& = & \Big( - \dfrac{(c_{L}^\dag)^{2} + (c_{R}^\dag)^{2}}{2} +  \dfrac{1}{d}  \sum_{k = 1 }^{d}  a_{k,\,L}^\dag b_{k,\,L}^\dag a_{k,\,R}^\dag b_{k,\,R}^\dag \Big)   |0\rangle  \nonumber\\
&=& - \sqrt{1-P} |\psi_f\rangle - \sqrt{P} |\gamma\rangle,
\end{eqnarray}
where
\begin{eqnarray} 
|\psi_f\rangle &=&  \frac{(c_{L}^\dag)^{2}+(c_{R}^\dag)^{2}}{2\sqrt{\chi_2}} |0\rangle, \\
|\gamma \rangle &=&  \frac{1}{\sqrt{d}}  \sum_{k = 1}^{d}  \Big( a_{k,L}^\dag  b_{k,L}^\dag a_{k,R}^\dag b_{k,R}^\dag \Big)   |0\rangle .
\end{eqnarray} 
We arrive at the desired state with probability $1-P$. Therefore, the probability of success depends on the degree of entanglement inside the coboson. In the limit of large entanglement ($d\gg 1$) the probability of success approaches one, since $P\rightarrow 0$. This reconfirms previous claims that the bosonic quality is related to the degree of entanglement.

%%%%%%%%%%%%%%%%%%%%%%%%%%%%%%%%%%%%%%%%%%%%%%%%

\section{Conclusions}

The above scenario is not a typical bunching scenario. In the standard case two elementary bosons bunch without an interaction. In case of cobosons an interaction is necessary to provide stability of the system for any evolution involving a BS. One may think that scenarios like \cite{e17} resemble the standard case, since the interaction between particles of type $a$ and $b$ seems only to provide stability of composite particles, whereas individual cobosons do not seem to interact. However, this is not true. Once both cobosons are in the same mode one is unable to distinguish which particle of type $a$ is interacting with which particle of type $b$. Therefore, in this case the interaction binds all the constituents together in a one big molecule which we interpret as a two-particle Fock state.

To conclude, we showed that an entangled spatially separated pair of fermions can exhibit some bosonic property. Namely, given two such pairs it is possible to locally produce a state which can be interpreted as an analog of a two-partite bosonic Fock state. However, a single fermionic pair cannot undergo a particle-like BS transformation without interaction between the spatially separated parts. Therefore, an entangled fermionic pair cannot be considered as a boson in an unambiguous way if there is no interaction between the constituents. Moreover, one has to be aware of the fact, that even if spatially separated fermions behaved like a single boson, such a bosonic particle would be very fragile since it would be prone to environmental disturbance \cite{Chudzicki}. 

{\it Acknowledgements.} Z.L. and P.K. were supported by the National Science Centre in Poland through the NCN Grant No. 2014/14/E/ST2/00585. D.K. was supported by the National Research Foundation and Ministry of Education in Singapore.

%%%%%%%%%%%%%%%%%%%%%%%%%%%%%%%%%%%%%%%%%%%%%%%%

%%%%%%%%%%%%%%%%%%%%%%%%%%%%%%%%%%%%%%%%%%%%%%%%


\begin{thebibliography}{99}

\bibitem{HOM}
C. K. Hong, Z. Y. Ou, and L. Mandel, Phys. Rev. Lett. {\bf 59}, 2044
(1987).

\bibitem{Law}
C. K. Law, Phys. Rev. A. {\bf 71}, 034306 (2005).

\bibitem{Chudzicki}
C. Chudzicki, O. Oke, and W. K. Wootters, Phys. Rev. Lett. {\bf 104}, 070402 (2010).

\bibitem{e2}
Y. H. Pong and C. K. Law, Phys. Rev. A 75, 043613 (2007).

\bibitem{e3}
M. Combescot, F. Dubin, andM. A. Dupertuis, Phys. Rev. A 80,
013612 (2009).

\bibitem{e4}
M. Combescot and O. Betbeder-Matibet, Phys. Rev. Lett. 104,
206404 (2010).

\bibitem{e16}
T. Brougham, S. M. Barnett, and I. Jex, J. Mod. Opt. 57, 587 (2010).

\bibitem{e5}
M. Combescot, S.-Y. Shiau, and Y.-C. Chang, Phys. Rev. Lett.
106, 206403 (2011).

\bibitem{e6}
M. Combescot, Europhys. Lett. 96, 60002 (2011).

\bibitem{e7}
R. Ramanathan, P. Kurzy\'{n}ski, T. K. Chuan, M. F. Santos, and D. Kaszlikowski, Phys. Rev. A 84, 034304 (2011).

\bibitem{e10}
A. Gavrilik and Y.Mishchenko, Phys. Lett. A 376, 1596 (2012).

\bibitem{e12}
M. C. Tichy, P. A. Bouvrie, and K. M\o{}lmer, Phys. Rev. A 86, 042317 (2012).

\bibitem{e17}
M. C. Tichy, P. A. Bouvrie, and K. M\o{}lmer, Phys. Rev. Lett. 109, 260403 (2012).

\bibitem{e18}
P. Kurzy\'{n}ski, R. Ramanathan, A. Soeda, T. K. Chuan, and D. Kaszlikowski, New J. Phys. 14, 093047 (2012).

\bibitem{e19}
A. Thilagam, J. Math. Chem. 51, 1897 (2013).

\bibitem{e11}
A. M. Gavrilik and Y. A. Mishchenko, J. Phys. A: Math. Theor. 46, 145301 (2013).

\bibitem{e8}
S.-Y. Lee, J. Thompson, P. Kurzy\'{n}ski, A. Soeda, and D. Kaszlikowski, Phys. Rev. A 88, 063602 (2013).

\bibitem{e13}
M. C. Tichy, P. A. Bouvrie, and K. M\o{}lmer, Phys. Rev. A 88, 061602 (2013).

\bibitem{e14}
M. C. Tichy, P. A. Bouvrie, and K. M\o{}lmer, Appl. Phys. B 117, 785 (2014).

\bibitem{e9}
S.-Y. Lee, J. Thompson, S. Raeisi, P. Kurzy\'{n}ski, and D. Kaszlikowski, New J. Phys. 17, 113015 (2015).

\bibitem{e15}
M. Combescot, R. Combescot, M. Alloing, and F. Dubin, Phys. Rev. Lett. 114, 090401 (2015).

\bibitem{e20}
A. Thilagam, Physica B: Condensed Matter 457, 232 (2015).

\bibitem{e21}
M. Combescot, S.-Y. Shiau, and Y.-C. Chang, Phys. Rev. A 93,
013624 (2016).

\bibitem{e22}
P. A. Bouvrie, M. C. Tichy, and K. M\o{}lmer, Phys. Rev. A 94, 053624 (2016)

%\bibitem{1art1}
%M. Fierz, Helv. Phys. Acta {\bf 12}, 3 (1939).

%\bibitem{2art1}
%W. Pauli, Phys. Rev. {\bf 58}, 716 (1940).

%\bibitem{3art1}
%A. Jabs, Found. Phys.  {\bf 40}, 776 (2009).

%\bibitem{1art2}
%S. S. Avancini, J. R. Marinelli, and G. Krein, J. Phys. A {\bf 36}, 9045 (2003).


\bibitem{book}
M. A. Nielsen, and I. L. Chuang, {\it Quantum Computation and Quantum Information}, Cambridge: Cambridge University Press (2000).

\bibitem{Yang}
C. N. Yang, Rev. Mod. Phys. {\bf 34}, 694 (1962).

\bibitem{Combescot}
M. Combescot, O. Betbeder-Matibet, and F. Dubin, Phys.
Rep. {\bf 463}, 215 (2008).

\bibitem{Eckert}
K. Eckert , J. Schliemann, D. Bruß, M.Lewenstein, Ann. Phys. {\bf 299}, 88 (2002).

\bibitem{Bell}
N. Brunner, D. Cavalcanti, S. Pironio, V. Scarani, S. Wehner, Rev. Mod. Phys. {\bf 86}, 419 (2014).


\end{thebibliography}
\end{document}